\documentclass[preprint2,times]{aastex631}
\shorttitle{The ambiguous nature of kinematic structures  in HD\,100546}
\shortauthors{S. Casassus et al.}
\graphicspath{{./}{figures/}}

\begin{document}


\title{The Doppler-flip in  HD\,100546 as a disk eruption: the elephant in the room of kinematic protoplanet searches}

\author[0000-0002-0433-9840]{Simon Casassus}
\affiliation{Departamento de Astronom\'{\i}a, Universidad de Chile, Casilla 36-D, Santiago, Chile}
\affiliation{Facultad de Ingenier\'ia y Ciencias, Universidad Adolfo Ib\'a\~nez, Av. Diagonal las Torres 2640, Pe\~{n}alol\'{e}n, Santiago, Chile }
\affiliation{Data Observatory Foundation, Chile}
\affiliation{Millennium Nucleus on Young Exoplanets and their Moons - YEMS, Chile.}

\author[0000-0003-0564-8167]{Miguel C\'arcamo}
\affiliation{Jodrell Bank Centre for Astrophysics, Department of Physics and Astronomy, University of Manchester,\\ ~~Alan Turing Building, Oxford Road, Manchester, M13 9PL, UK}
\affiliation{University of Santiago of Chile (USACH), Faculty of Engineering, Computer Engineering Department, Chile}
\affiliation{Center for Interdisciplinary Research in Astrophysics and Space Exploration (CIRAS), Universidad de Santiago de Chile}

\author[0000-0001-5073-2849]{Antonio Hales}
\affiliation{Joint ALMA Observatory, Avenida Alonso de C\'ordova 3107, Vitacura 7630355, Santiago, Chile}
\affiliation{National Radio Astronomy Observatory, 520 Edgemont Road, Charlottesville, VA 22903-2475, United States of America}

\author[0000-0002-3354-6654]{Philipp Weber}
\affiliation{Departamento de Física, Universidad de Santiago de Chile, Av. Victor Jara 3493, Estación Central, Santiago, Chile}
\affiliation{Millennium Nucleus on Young Exoplanets and their Moons - YEMS, Chile.}
\affiliation{Center for Interdisciplinary Research in Astrophysics and Space Exploration (CIRAS), Universidad de Santiago de Chile}

\author[0000-0002-2490-1079]{Bill Dent}
\affiliation{Joint ALMA Observatory, Avenida Alonso de C\'ordova 3107, Vitacura 7630355, Santiago, Chile}
\affiliation{European Southern Observatory, Avenida Alonso de C\'{o}rdova 3107, Vitacura 7630355, Santiago, Chile}





\begin{abstract}
  The interpretation of molecular-line data in terms of hydro
  dynamical simulations of planet-disk interactions fosters new hopes
  for the indirect detection of protoplanets. In a model-independent
  approach, embedded protoplanets should be found at the roots of
  abrupt Doppler flips in velocity centroid maps. However, the largest
  velocity perturbation known for an unwarped disk, in the disk of HD\,100546,
  leads to a conspicuous Doppler flip that coincides with a thick dust
  ring, in contradiction with an interpretation in terms of a
  $\gtrsim 1 \,M_{\rm jup}$ body.  Here we present new ALMA
  observations of the $^{12}$CO(2-1) kinematics in HD\,100546, with a
  factor of two finer angular resolutions. We find that the disk
  rotation curve is consistent with a central mass
  $ 2.1 < M_\star /M_\odot < 2.3$, and that the blue-shifted side of
  the Doppler flip is due to vertical motions, reminiscent of the disk
  wind proposed previously from blue-shifted SO lines. We tentatively
  propose a qualitative interpretation in terms of a surface
  disturbance to the Keplerian flow, i.e. a disk eruption, driven by
  an embedded outflow launched by a $\sim 10\,M_{\rm earth}$
  body. Another interpretation involves a disk-mass-loading hot-spot
  at the convergence of an envelope accretion streamer.
  
\end{abstract}



\section{Introduction} \label{sec:intro}

The first exoplanets were detected indirectly through the stellar
reflex motion \citep[][]{MayorQueloz1995Natur.378..355M}. Likewise,
the gas velocity field in protoplanetary disks bears the imprint of
planet/disk interactions, and molecular-line observations have emerged
as an alternative approach to the identification of protoplanets. The
first observability predictions, by \citet{Perez2015ApJ...811L...5P},
linked embedded planets with kinks or wiggle-shaped deviations from
sub-Keplerian rotation in channel maps (sky images in
position-velocity datacubes, hereafter $I_v(\vec{x})$). These
predictions appear to match the data in HD\,163296 and in HD\,97048,
where the most conspicuous deviations are interpreted in terms of the
location and mass of the perturber through comparison with hydro
simulations \citep{Pinte2018ApJ...860L..13P,
  Pinte2019NatAs...3.1109P}.


New hopes for protoplanet detection are thus coming from the disk
kinematics, with the interpretation of molecular-line data in terms of
hydro dynamical simulations. However, the simulations required to link
the channel-map data with embedded bodies tend to idealise the
systems. The unicity of a given planetary system configuration remains
to be explored.  It turns out that wiggles or kinks are almost
ubiquitous within a disk, and also very frequently observed, as they
are picked up in at least nine disks
\citep[][]{Pinte2020ApJ...890L...9P} from the DSHARP ALMA Large
Program \citep[][]{Andrews2018ApJ...869L..41A}, despite these
observations not being designed to pick-up small velocity
structures. It may be that the ubiquity of velocity deviations
reflects the global impact of gravitational interactions with the
embedded planetary systems, or even large-scale perturbations in
gravitational unstable disks \citep[][]{Hall2020,
  Paneque2021}. Another uncertainty is that a fraction of these
structures stems from synthesis imaging limitations, or from structure
in the underlying optical depth. In any case, although velocity
structures are expected from planet/disk interactions, it is not
possible to unequivocally infer the location of embedded planets
directly from the observed kinks or wiggles.

An alternative to pick-up and locate embedded protoplanets has been
proposed based on the velocity centroid maps,
$v_\circ(\vec{x}) = \langle v\, I_v \rangle = \int dv\, v \,I_v/ \int I_v
dv$.  After subtraction of the axially symmetric background flow, the
velocity reversal along the planetary wakes, right at the location of
the planet, should be observable as a Doppler-flip in molecular line
maps \citep{Perez2018MNRAS.480L..12P,
  CasassusPerez2019ApJ...883L..41C}, i.e. as an abrupt and localised
change of sign in the velocity centroid of molecular line maps.

Very deep and long-baseline ALMA observations of HD\,100546 \citep[up
to 12\,km,][]{Perez2020ApJ...889L..24P} revealed the largest known
velocity deviations seen in channel maps. After subtraction of the
background axially symmetric flow, these deviations translate into a
conspicuous Doppler flip, whose amplitude reaches 1/3 of the Keplerian
velocity \citep{CasassusPerez2019ApJ...883L..41C}.  The root of the
Doppler flip is constrained within $ \sim$5\,au, at stellocentric
radius $R_f \sim 28 \pm 3$\,au. The amplitude of the flip requires a
rather massive body, $\gtrsim 5\,M_{\rm jup}$ for an interpretation in
terms of an embedded planet. However, as noted by
\citet{CasassusPerez2019ApJ...883L..41C}, its coincidence with the
massive dust ring of HD\,100546's disk  is in contradiction with a massive
proto-giant, as theory predicts the clearing of a gap by the
protoplanet. The only Doppler flip reported in the literature,
corresponding to the largest velocity deviations known, is thus the
elephant in the room of kinematic protoplanet searches. It is
pointless to search for planets causing small velocity deviations, if
the largest velocity deviation known remains unexplained.

In this letter we present new ALMA observations of HD\,100546 in
$^{12}$CO(2-1) and adjacent continuum, at unprecedented angular
resolutions (Sec.\,\ref{sec:obs}). We analyze the line kinematics in
terms of 3-D velocity deviations from an axially symmetric flow
(Sec.\,\ref{sec:kin}), and discuss these results in terms of possible
origins for the kinematic structures (Sec.\,\ref{sec:discuss}), before
summarizing our results (Sec.\,\ref{sec:conc}).

\section{Observations} \label{sec:obs}

\begin{deluxetable*}{ccccccccc}
\tablecaption{Summary of ALMA Observations \label{log1}}
\tablewidth{700pt}
\tabletypesize{\scriptsize}
\tablehead{
\colhead{Date} & 
\colhead{N Ant.} & 
\colhead{Execution Block} & 
\colhead{ToS } & 
\colhead{Avg. Elev. } & 
\colhead{Mean PWV } & 
\colhead{Phase RMS } & 
\colhead{Baseline } & 
\colhead{MRS} \\
\colhead{} & 
\colhead{} & 
\colhead{} & 
\colhead{(sec)} & 
\colhead{(deg)} & 
\colhead{(mm)} & 
\colhead{(deg)} & 
\colhead{(m)} & 
\colhead{(")} 
}
\startdata
2019-06-04  & 45&uid://A002/Xdd3de2/X22e  & 5468&  42.6 & 1.1 & 7.6 & 83  - 15238 & 0.4 \\ 
2019-06-04  & 45&uid://A002/Xdd3de2/Xa95  & 5547&  41.5 & 1.1 & 8.0 & 83  - 15238 & 0.4 \\
2019-06-05  & 45&uid://A002/Xdd3de2/X147a & 5584&  36.0 & 1.1 & 9.5 & 83  - 15238 & 0.4 \\
2019-06-05  & 41&uid://A002/Xdd4cf3/X282  & 5550&  42.3 & 0.9 & 8.8 & 236 - 15238 & 0.3 \\
2019-06-06  & 41&uid://A002/Xdd4cf3/Xe0c  & 5561&  37.9 & 0.9 & 9.0 & 236 - 15238 & 0.3 \\
2019-06-21  & 45&uid://A002/Xddf4b5/X5b5  & 5584&  41.8 & 1.3 & 12.2& 83  - 16196 & 0.3 \\
\enddata

\tablecomments{Summary of the new ALMA observations presented in this
  work. The table shows the date of the observations, the total number
  of antennas in each executions, the total time on source (ToS),
  target average elevation, mean precipitable water vapor column (PWV)
  in the atmosphere, antenna-based phase noise, minimum and maximum
  baselines, and maximum recoverable scale (MRS) for the used array
  configuration.} \label{table:log}
\end{deluxetable*}

The data presented in this letter correspond to ALMA observations of
HD\,100546, carried out in program 2018.1.01309.S
(P.I. S.\,P\'erez).
The source was acquired in 6 executions
blocks, for a total of 9.2 hours on-source.  The array was in
configuration C-10, with baselines ranging from 83 to 15238 meters, and 41-45 active antennas.
A log of observations is provided in
Table\,\ref{table:log}. Quasars J1107-4449 and J1145-6954 were used for flux and phase calibration, respectively.

The correlator was setup to provide 4 spectral windows (spws). The
continuum was sampled with spw\,0, 1 and 2, at 233.0\,GHz, 217.0\,GHz
and 219.0\,GHz, each covering 2\,GHz and thus yielding a total
bandwidth of 6\,GHz. The $^{12}$CO(2-1) line, at a rest frequency of
230.538\,GHz, was sampled in spw\,3 with 158.74\,m\,s$^{-1}$ channels,
which we resampled to the local standard of rest into 160\,m\,s$^{-1}$
channels. The data were calibrated by staff from the North America ALMA Regional Center.

\begin{figure*}
  \centering
  \includegraphics[width=\textwidth,height=!]{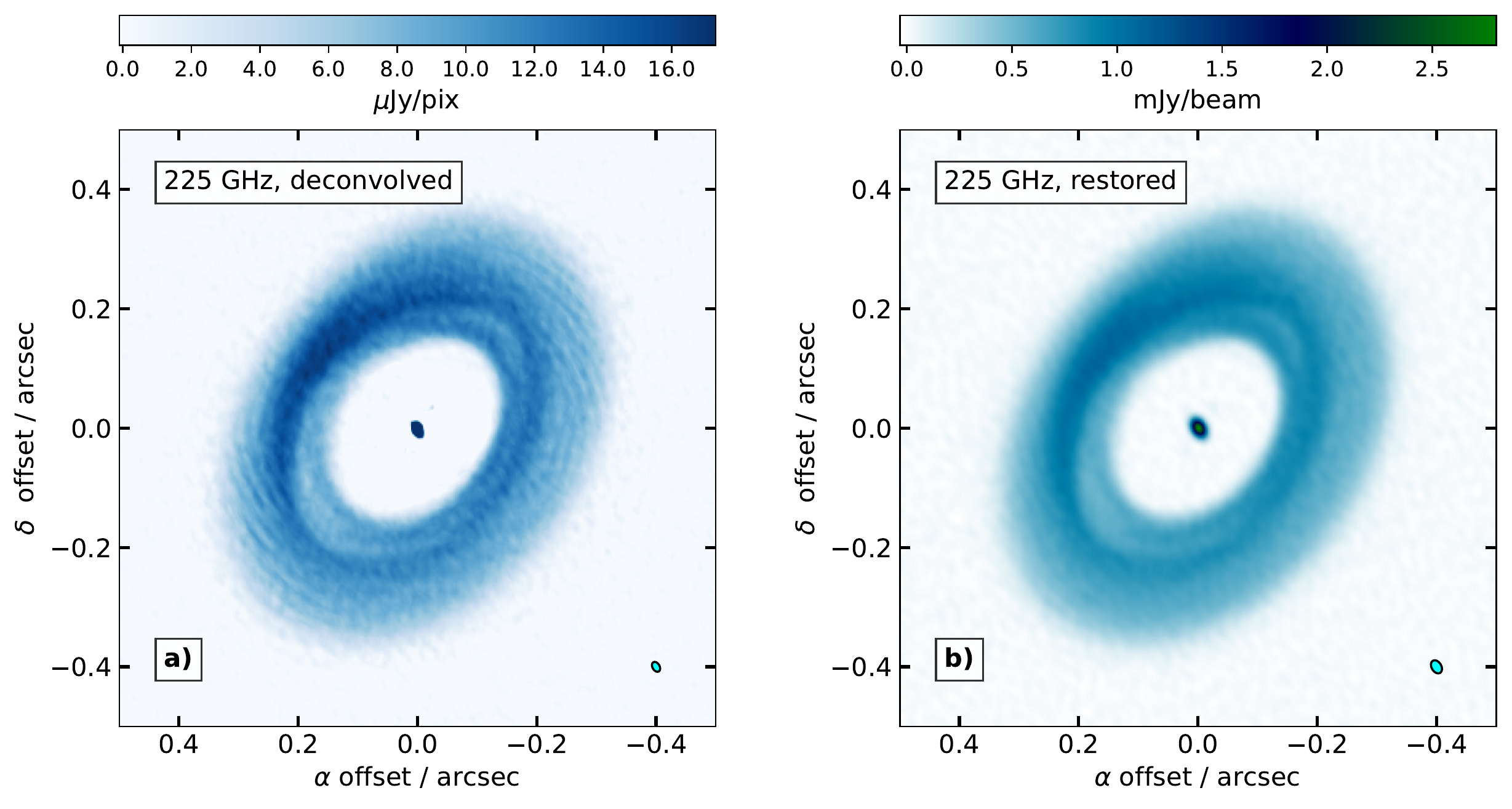}
  \caption{Continuum at 225\,GHz. {\bf a)}: model image $I^m$ obtained with
    {\tt uvmem}, with an effective beam $0\farcs018\times0\farcs012$
    at 26\,deg (as estimated with a elliptical Gaussian fit to the
    central source, which appears unresolved). {\bf b)}: restored
    image, with a Briggs robustness parameter of 0, corresponding to a
    beam $0\farcs024\times 0\farcs016$ arcsec at PA $31.1$. The noise in the
    image is 9\,$\mu$Jy\,beam$^{-1}$.} \label{fig:cont}
\end{figure*}

In this report we focus on the disk kinematics and continuum emission
of the central regions of the HD\,100546 disk, along its inner ring,
which is only $\sim0\farcs2$ in radius. We report exclusively on the
new long baseline data, and do not concatenate with the more compact
configuration previously presented in
\citet[][]{{Perez2020ApJ...889L..24P}}.  Keplerian rotation spans
35\,mas in the time interval of $\sim$2\,yrs between the acquisitions
of the present data and that of the intermediate-length
configuration. This arc is almost 3$\times$ the finest angular scales
reported here \citep[about $0\farcs12$ or even $0\farcs09$, as measured in
the central source assuming that it is unresolved, see
Fig.\ref{fig:cont}, or as estimated with 1/3 the natural-weights beam,
see][]{Carcamo2018A&C....22...16C} and almost twice those from
\citet[][]{{Perez2020ApJ...889L..24P}}. The concatenated continuum
image of the ring is thus smoothed to angular scales 2 to 3 times
coarser. However, the concatenated $^{12}$CO(2-1) data is much more
sensitive to extended emission. Its presentation is postponed to a
forthcoming article on the large scale disk kinematics.




The data were self-calibrated using the standard procedure. Each
execution block (EB) was self-calibrated independently. We set up CASA
task {\tt gaincal} to average whole scans (option `solint' set to
`inf'). The self-calibration process provided an improvement of a
factor of 1.5-2 per EB in {\tt tclean} images using Briggs weights
with robustness parameter $r=0.0$, with typical dynamic ranges
(i.e. signal-to-noise ratio) of $\sim$ 175 to $\sim$ 200.

Following the self-calibration procedure, we performed the final
synthesis imaging with the {\sc uvmem} package
\citep[][]{Casassus2006, Carcamo2018A&C....22...16C}, in the same way
as in \citet[][]{Casassus2021MNRAS.507.3789C}. In the present
application, {\sc uvmem} produces a positive-definite model-image
$I^m$ and corresponding model visibilities $V^m_k$ that fit the visibility data
$V^\circ_k$ in a least-square sense, i.e. by minimizing:
\begin{equation}
  \chi^2 = \sum_{k=1}^{N_{\rm vis}} \omega_k  |V^\circ_k - V^m_k|^2
  \label{eq:L},
\end{equation}
where $\omega_k$ correspond to the visibility weights and where the
sum runs over all visibility data (i.e. without gridding).

For the restoration of the continuum image, shown in
Fig.\,\ref{fig:cont}, we use Briggs weights with robustness parameter
$r=0$, resulting in a clean beam $0\farcs024\times 0\farcs016 / 31.1$, where we
give the beam major axis (bmaj), minor axis (bmin) and direction (bpa)
in the format bmaj$\times$bmin/bpa. This restored image can be
compared to that obtained by standard packages such as {\tt tclean} in
{\sc casa}.

The channel maps of the $^{12}$CO(2-1) line, shown in
Appendix\,\ref{sec:maps}, were restored with Briggs $r=1$, for a beam
$0\farcs038\times 0\farcs024 / 0.0 $. In the case of emission that is much more
extended than the beam, the level of aliasing is reduced with masks
that isolate the signal. Here we used the same strategy as in
\citet[][]{Casassus2021MNRAS.507.3789C}, with Keplerian masks obtained
with the
tool\footnote{\url{https://github.com/richteague/keplerian_mask}}
developed by \citet{rich_teague_2020_4321137}.

\begin{figure*}
  \centering
  \includegraphics[width=\textwidth,height=!]{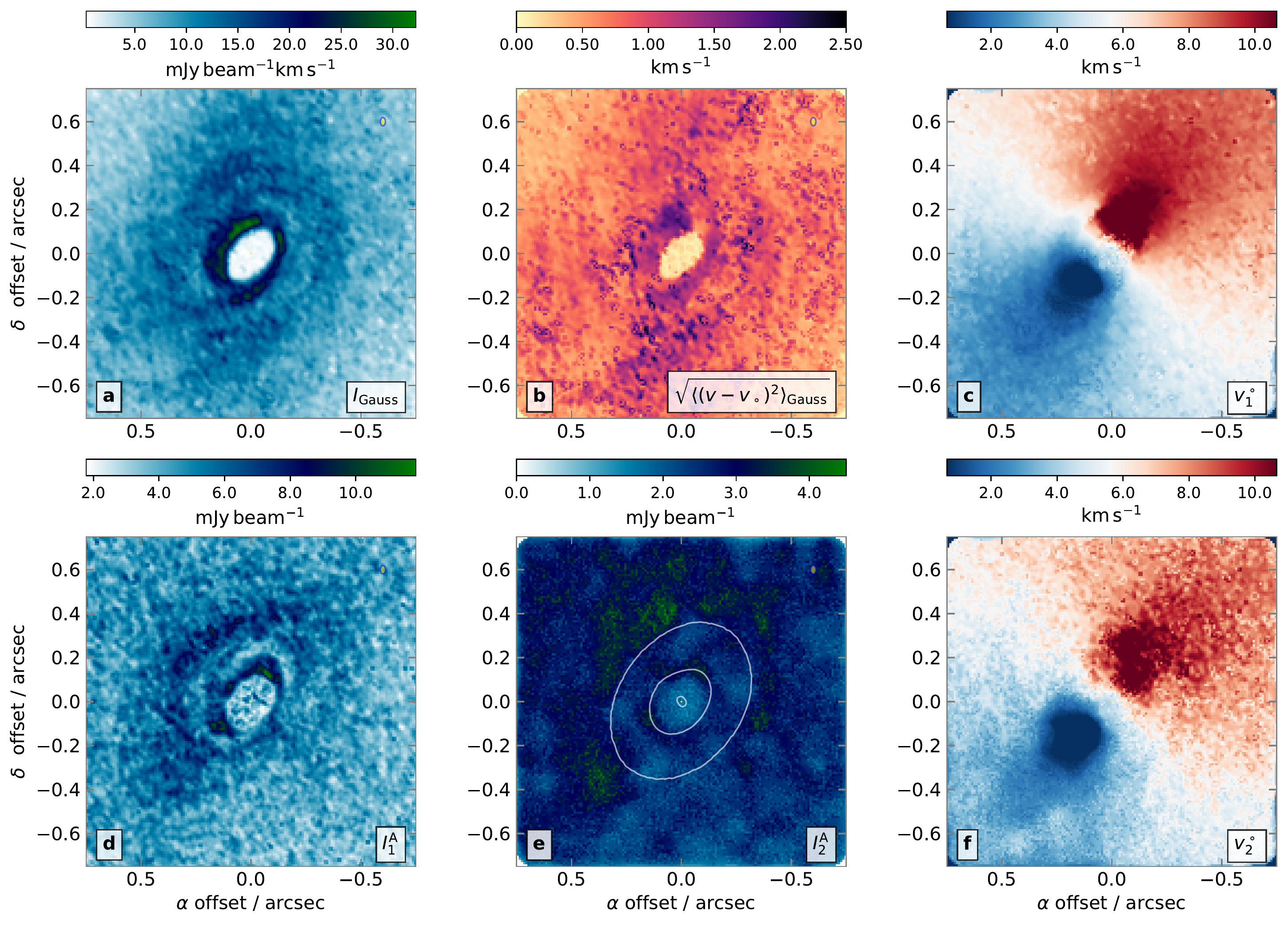}
  \caption{Moment maps in $^{12}$CO(2-1) using double-Gaussian fits in
    velocity.  {\bf a)} The 2-Gaussian velocity-integrated
    intensity. {\bf b)} The velocity dispersion extracted using the
    two Gaussians (this image has been median-filtered).  {\bf c}) The
    velocity centroid of the brighter Gaussian. {\bf d)} The amplitude
    of the brighter Gaussian, $I_1^A$. {\bf e}, {\bf f:} same as {\bf
      c} and {\bf d}, but for the fainter Gaussian (with amplitude
    $I_2^A$, these images have been median-filtered). In e) the
    continuum ring is overlaid against $I_2^A$ in a single contour at
    1/20 peak, and  is roughly outlined as a region of
    fainter  amplitude.} \label{fig:moments}
\end{figure*}

Moments maps were extracted with double-Gaussian fits to each line of
sight, using package {\sc
  GMoments}\footnote{\url{https://github.com/simoncasassus/GMoments}}
\citep[][]{Casassus2021MNRAS.507.3789C}. The line profile for a given
line of sight is modeled as
$I_v = \sum_{j=1}^2 I^A_j
\exp\left(-\frac{\left(v-v^\circ_j\right)^2}{\sigma_j^2}\right)$. The
top side, that faces the observer, is approximately traced with the
brighter Gaussian, with largest amplitude.  The moment maps that
result from the double-Gaussian fits are summarized in
Fig.\,\ref{fig:moments}.  The following analysis of disk kinematics is
based on Fig.\,\ref{fig:moments}c. Systemic velocity was set to
5.71\,km\,s$^{-1}$, based on the kinematic analysis presented below
(Sec.\,\ref{sec:kin}).


\section{Disk rotation curve and kinematics } \label{sec:kin}

Disk rotation can be approximated with an axially-symmetric velocity
field $\vec{\tilde{v}}(\vec{r})$ in terms of a 3-D disk rotation curve
$\vec{\tilde{v}}(R)$, with
$\vec{\tilde{v}}(\vec{r}) =
(\tilde{v}_R(R),\tilde{v}_\phi(R),\tilde{v}_z(R))$ in cylindrical
coordinates.  The deviations from the axially-symmetric flow are then
$\vec{u}(\vec{r}) = \vec{v}(\vec{r}) - \vec{\tilde{v}}(\vec{r})$,
where $\vec{v}(\vec{r})$ is the 3-D velocity field of the gas. If the
emission stems from a thin layer at the unit-opacity surface, the
first moment of the axially symmetric flow along a line of sight
$\hat{s}(\vec{x})$ is
\begin{equation}
v^m_\circ(\vec{x}) = \hat{s}(\vec{x}) \cdot \vec{\tilde{v}}\left(\vec{f}_{\mathrm{PA},i,\psi}^{-1}(\vec{x})\right),
\end{equation}
where $\vec{x} = \vec{f}_{\mathrm{PA},i,\psi}(R,\phi)$. The coordinate
transform $\vec{f}$ depends on disk position angle PA and inclination
$i$, and relates $\vec{x}$ to the point, seen in the disk's
cylindrical system, at the intersection of the line of sight with the
disk surface, which is approximated locally as a cone with opening
angle $\psi$. The observed velocity deviations are thus $\delta
v_\circ (\vec{x}) = v_\circ - v^m_\circ$, and derive from the
intrinsic velocity deviations $\vec{u}(\vec{r})$.

Here we apply the {\sc ConeRot} package\footnote{
  \url{https://github.com/simoncasassus/ConeRot}}, described in
\citet{CasassusPerez2019ApJ...883L..41C, Casassus2021MNRAS.507.3789C},
to derive the 3-D rotation curve
$\vec{\tilde{v}}(R)=(\tilde{v}_R(R), \tilde{v}_\phi(R),
\tilde{v}_z(R))$, in disk-centred cylindrical coordinates. Details on the extraction of this 3-D rotation curve are given in Appendix\,\ref{sec:conerot}. We focus on the
vicinity of the Doppler flip, and postpone an analysis of the whole
disk kinematics to a forthcoming article. Although {\sc ConeRot} can
also estimate the disk orientation based on the line data alone, these
orientation estimates are partly degenerate with the rotation
curve. We therefore fixed the disk orientation to that obtained from
the continuum, which was extracted using using the {\sc MPolarMaps}
package
\citep[\url{git@github.com:simoncasassus/MPolarMaps.git},][]{Casassus2021MNRAS.507.3789C}.
This package minimizes the variance in the radial profile of the
continuum intensity to estimate PA and $i$. The result is
PA=$323\pm0.7$\,deg, $i=41.7\pm0.4$\,deg. The ring center was
estimated to be offset from the phase center (itself pointed at the
star), by $\Delta \alpha = 0\farcs008\pm 0\farcs0008 $ and
$\Delta \delta = 0\farcs001\pm0\farcs0007$.

\begin{figure}
  \centering
\includegraphics[width=\columnwidth,height=!]{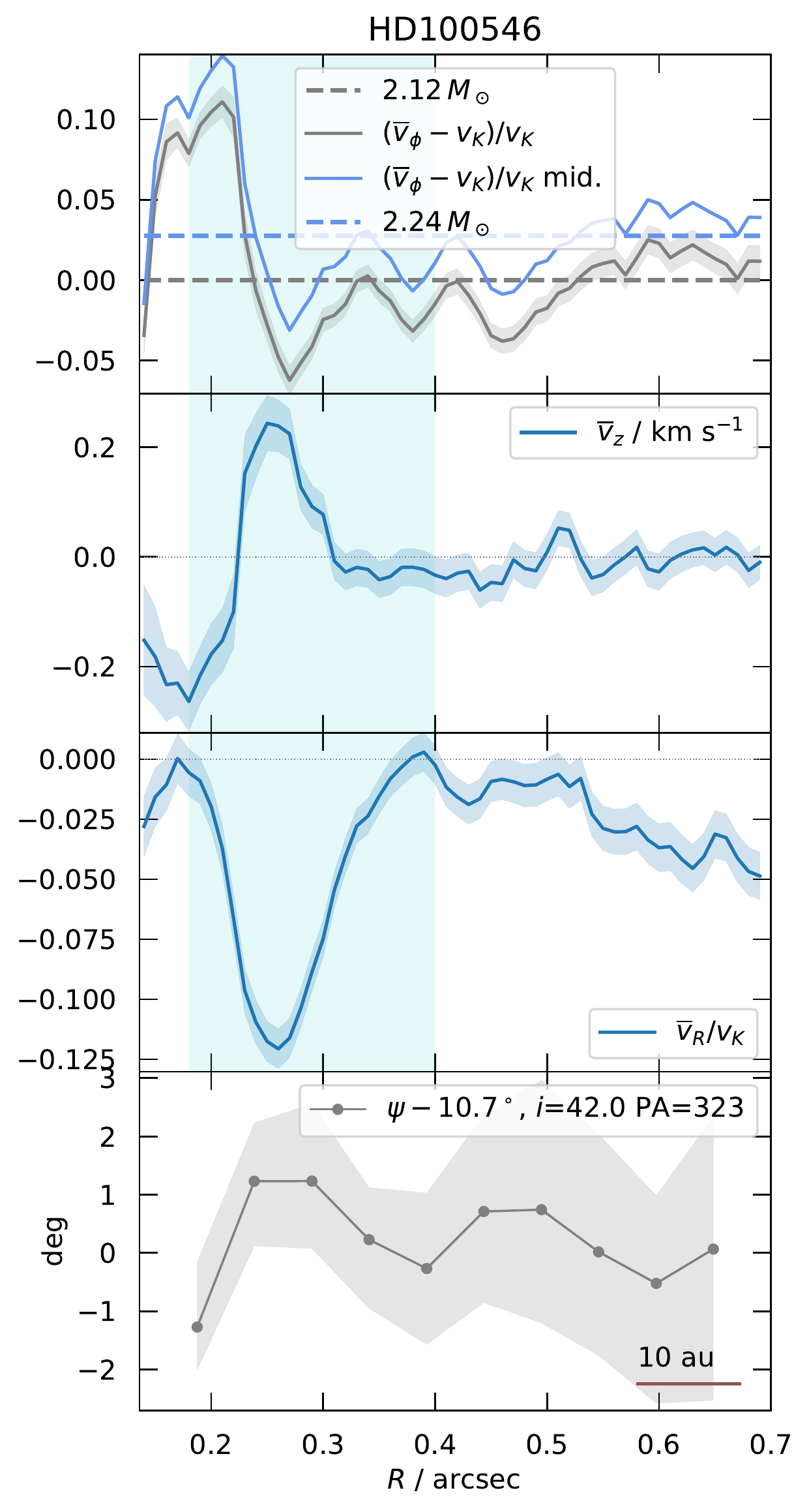}
\caption{Rotation curve in HD\,10054B.  The regions in cyan correspond
  to the total extent of the bright continuum ring. From top to
  bottom, we show: {\bf 1)} The azimuthal rotation curve
  $\tilde{v}_\phi(R)$. The dashed horizontal lines are comparison
  Keplerian profiles with the corresponding stellar mass.  The curve
  labelled `mid' is an extrapolation to the disk mid-plane assuming
  vertical Keplerian shear.  {\bf 2)}: The vertical velocity component
  curve $\tilde{v}_z(R)$, where $\tilde{v}_z >0$ points away from the
  disk mid-plane. {\bf 3)}: The radial velocity component
  $\tilde{v}_r(R)$, where $\tilde{v}_r >0$ points away from the star .
  {\bf 4)}: The opening angle of the cone tracing the unit opacity
  surface for $^{12}$CO(2-1).} \label{fig:rotcurve}
\end{figure}

The rotation curve for the HD\,100546 disk, with meridional flows, is
shown in Fig.\,\ref{fig:rotcurve}. The rotation curve for a purely
azimuthal flow is very similar to Fig.\,\ref{fig:rotcurve}a, although
with a somewhat higher stellar mass (
$2.13 \pm 0.04 < M_\star /M_\odot < 2.26 \pm 0.04$).  The range of
possible values for the stellar mass, including both cases with and
without meridional flows, is
$2.12 \pm 0.04 < M_\star /M_\odot < 2.26 \pm 0.04$, where the upper
limits stems from vertical Keplerian shear, while the lower limit
stems from cylindrical rotation (see
Appendix\,\ref{sec:conerot}). These mass estimates, which assume a
distance of $108.1\pm0.4$\,pc \citep[][]{Gaia2018A&A...616A...1G},
bracket the photospheric estimate of
$M_\star = 2.18^{+0.02 }_{-0.17}\,M_\odot$
\citep[][]{Wichittanakom2020MNRAS.493..234W}. The lower mass estimate
of $1.83\pm0.01\,M_\odot$ reported in
\citet{CasassusPerez2019ApJ...883L..41C} was affected by the
extraction of the moment map using a single Gaussian, which is more
contaminated by the back side of the disk than the two-Gaussian, and
which led to higher inclinations \citep[of $\sim 46\,$deg, the stellar
mass is sensitive to the $\sim 2^{\rm nd}$ power of
inclination,][their Eq.\,10]{Casassus2021MNRAS.507.3789C}, along with  a
missing $\sin(i)$ correction.


\begin{figure*}
  \centering
 \includegraphics[width=\textwidth,height=!]{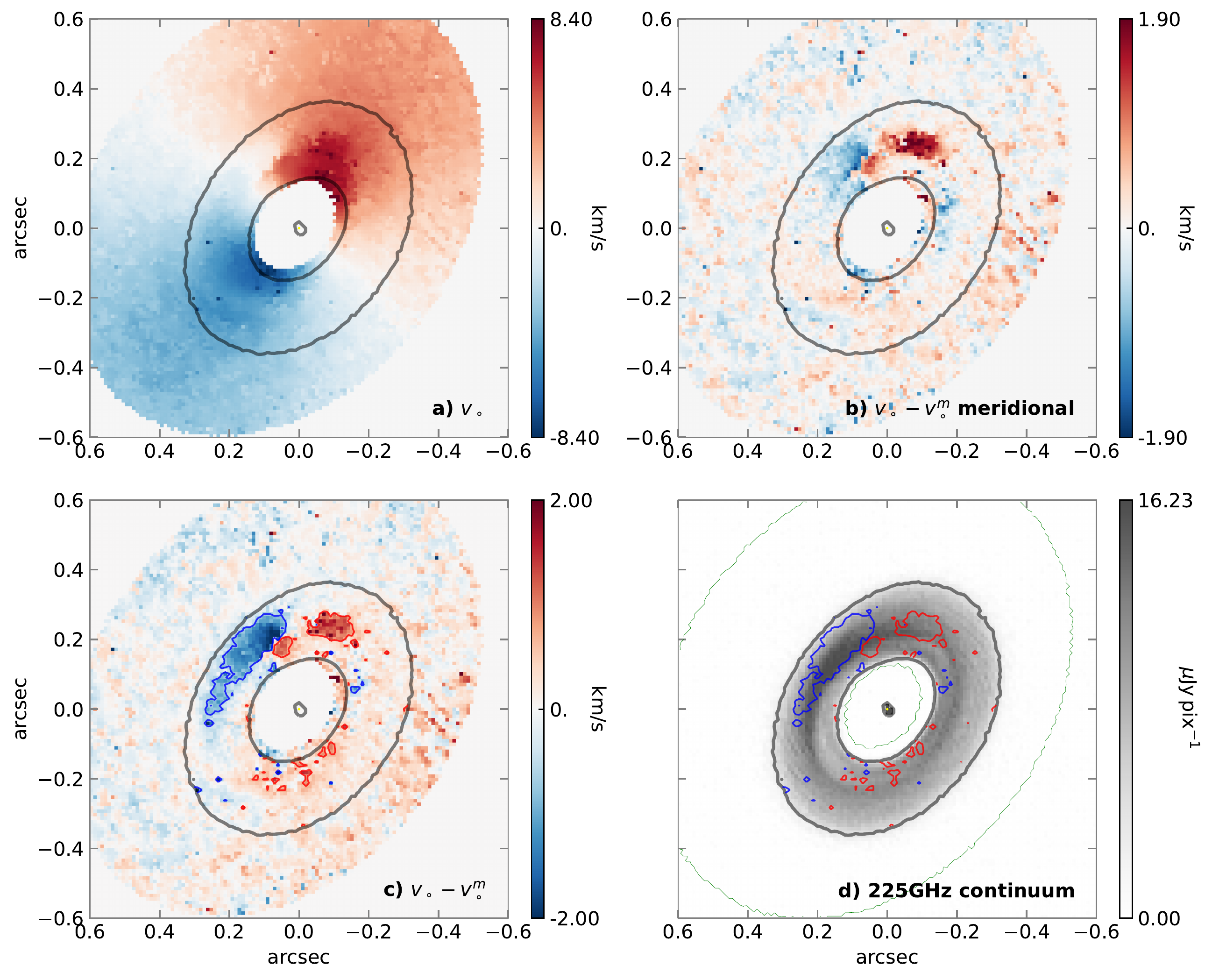}
 \caption{Non-axial kinematics in the disk of HD\,100546. {\bf a}: the velocity centroid $v_\circ$ from
   Fig.\,\ref{fig:moments}c. {\bf b}: $v_\circ$ after subtraction of
   the axially symmetric flow, $v_\circ^m$, including radial and
   vertical components (i.e. meridional flows) in the rotation
   curve. {\bf c}: $v_\circ - v_\circ^m$, same as b) but without
   meridional flows. The single red and blue contours correspond to
   $\pm 0.7\,$km\,s$^{-1}$.   {\bf d}: Continuum from
   Fig.\,\ref{fig:cont}a, with an overlay of the contours for
   $v_\circ - v_\circ^m$ from c). } \label{fig:kin}
 
\end{figure*}

\begin{figure*}
  \centering
 \includegraphics[width=\textwidth,height=!]{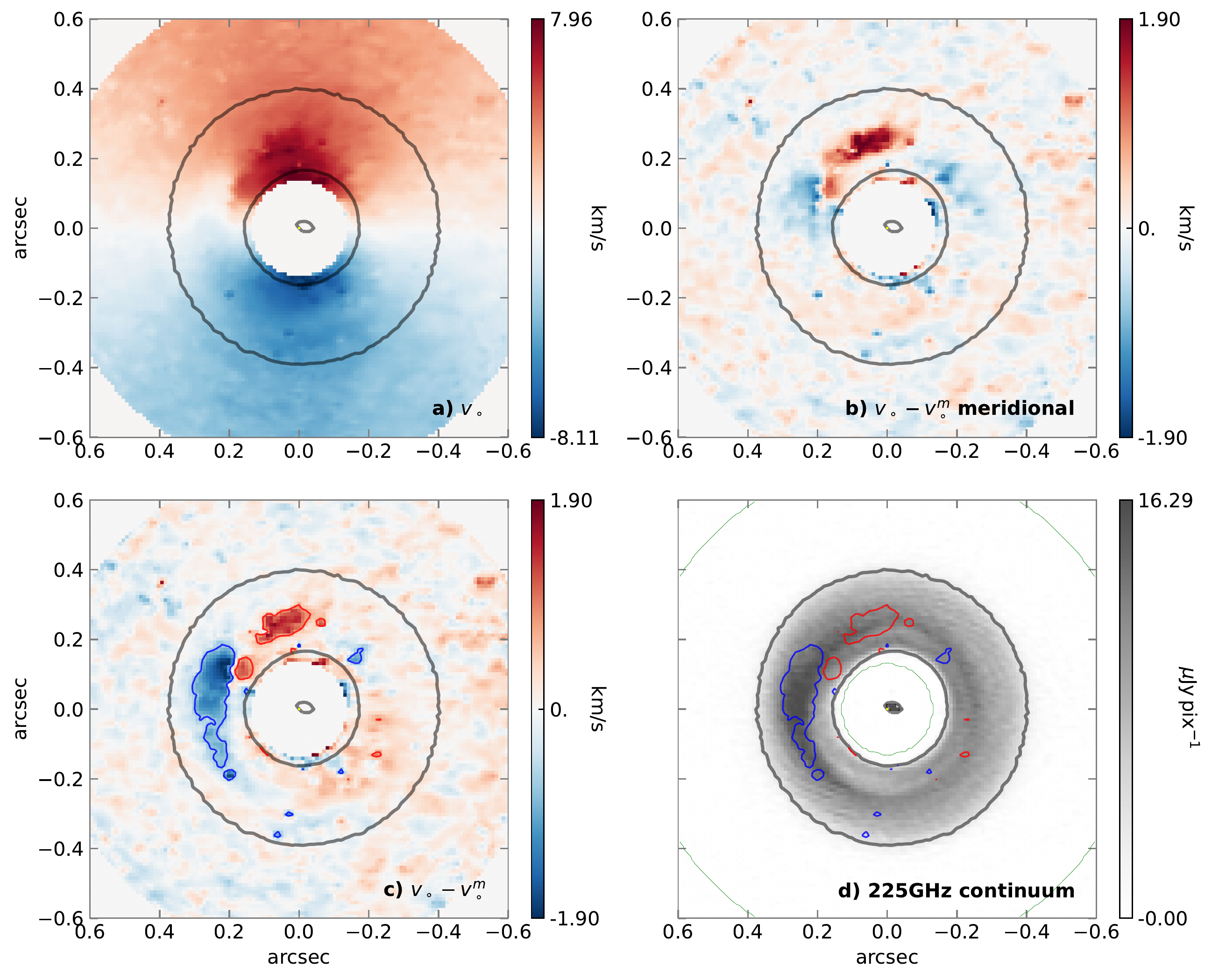}
 \caption{Face-on view of the non-axial kinematics in the disk of HD\,100546. Annotations follow from Fig.\,\ref{fig:kin}.} \label{fig:kinfaceon}
 
\end{figure*}

Given the disk rotation curve, we now turn to the non-axial
kinematics, i.e. to local deviations from axially symmetric
flow. Fig.\,\ref{fig:kin} presents a summary of the kinematics, as
seen on the sky, and compares them with the continuum emission. The
same kinematic structures are shown in a face-on deprojection in
Fig.\,\ref{fig:kinfaceon}. The de-projection technique accounts for
the thickness of the CO(2-1) surface, and is described in
\citet{CasassusPerez2019ApJ...883L..41C}.



\section{Discussion} \label{sec:discuss}

\begin{figure}
  \centering
  \includegraphics[width=\columnwidth,height=!]{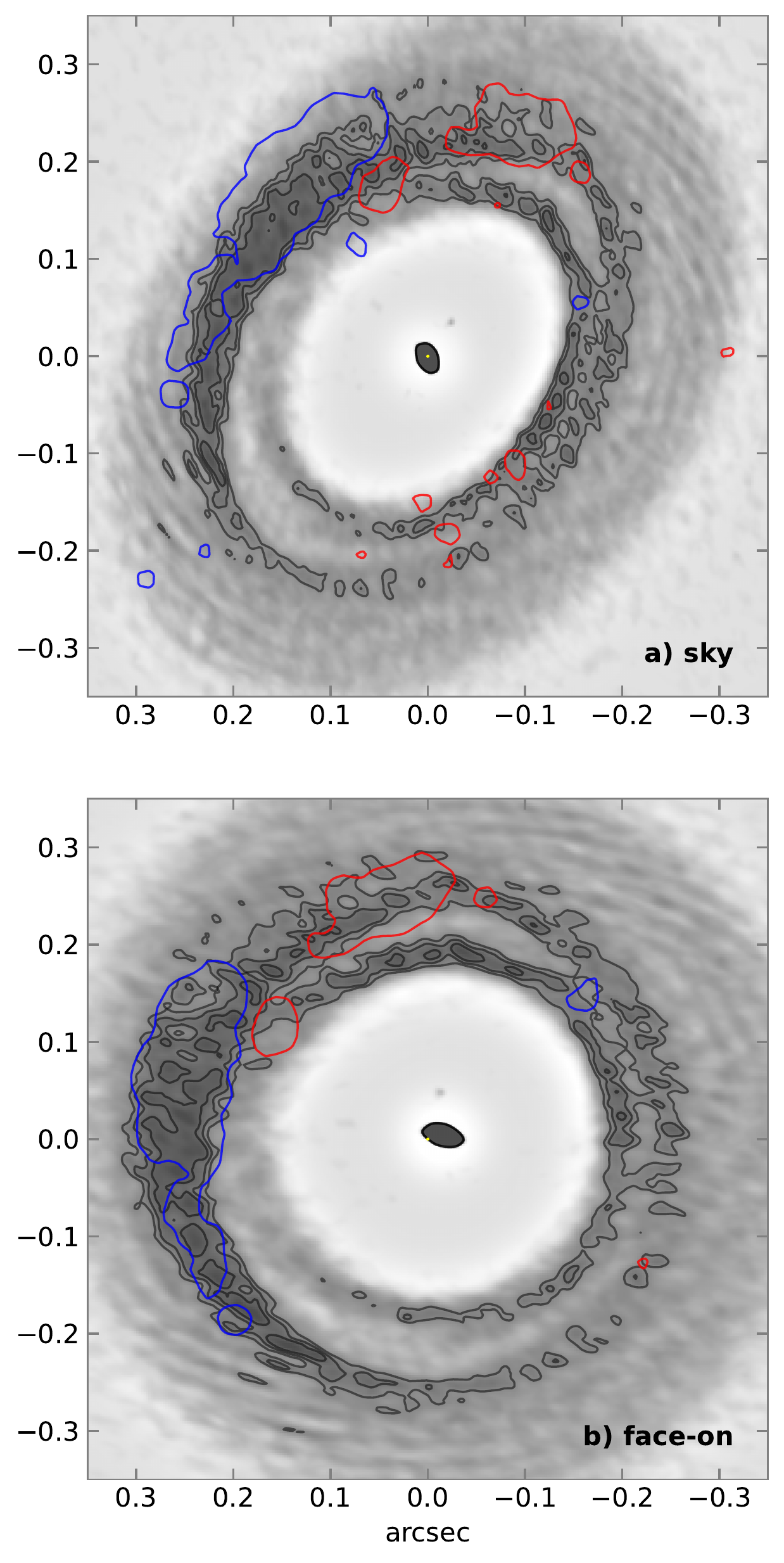}
\caption{{\bf a:} Sky view of the Doppler-flip in the disk of HD\,100546, as in
  Fig.\,\ref{fig:kin}d, but overlaid on a version of the {\tt uvmem}
  model image that has been processed with an unsharp mask \citep[as
  described in][]{Perez2020ApJ...889L..24P}. The grey contours, at
  6\%, 7\% and 8\% maximum, are chosen to highlight structure along
  the ring. {\bf b:} Same as a), but from a face-on
  perspective. \label{fig:unsharpflip}}
\end{figure}

The velocity deviations from a purely azimuthal flow, inferred from
Figs.\,\ref{fig:kin}c and \ref{fig:kinfaceon}, are very similar to
those reported in \citet[][]{CasassusPerez2019ApJ...883L..41C}.  There
is a coherent velocity structure in the form of an arc that appears to
coincide with the bright sub-mm continuum ring. The correspondence
with structures in the continuum is shown in more detail in
Fig.\,\ref{fig:unsharpflip}, where we also see that the center of the
flip coincides with a bifurcation along this ridge.

\begin{figure}
  \centering
\includegraphics[width=\columnwidth,height=!]{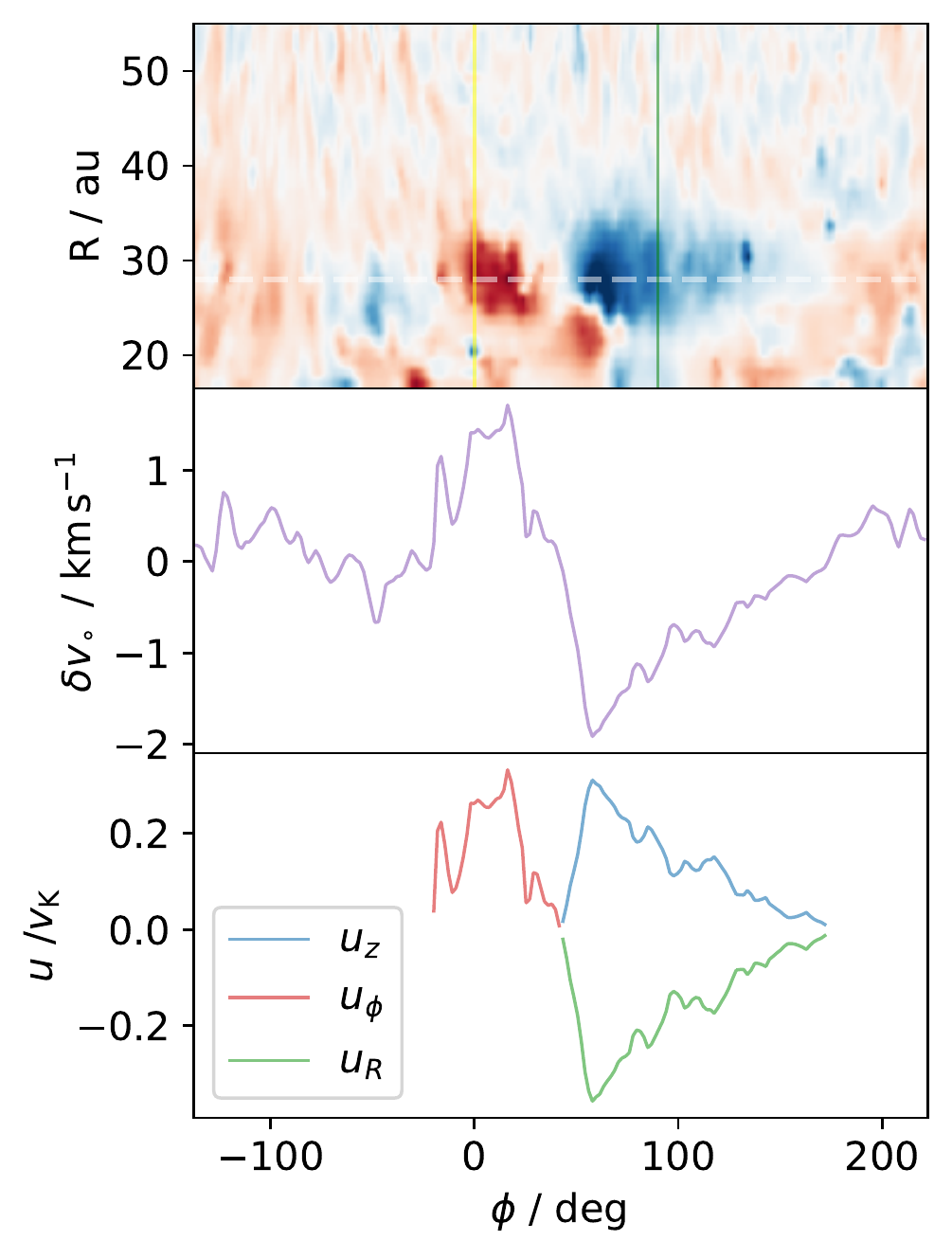}
\caption{{\bf Top:} Polar expansion of $\delta v_\circ$, from the
  face-on deprojection in Fig.\,\ref{fig:kinfaceon}, and with a 1D
  rotation curve (no meridional flows). The horizontal dashed line
  corresponds to $R_f=28\,$au. The vertical yellow and green lines
  each indicate the directions of the disk major and minor axis,
  respectively.  {\bf Middle:} extraction of $\delta v_\circ$ at
  $R_f=29\,$au,
  $\delta
  v_\circ(\vec{x}=\vec{f}_{\mathrm{PA},i,\psi}(R_f,\phi))$. {\bf
    Bottom:} Intrinsic velocity field $\vec{u}$, assuming that the red
  side of the flip is entirely due to azimuthal velocity deviations,
  while the blue side is due to either radial or vertical components,
  such that $u_z>0$ corresponds to a wind, and $u_R< 0$ to
  stellocentric accretion.  The $x-$axis of all plots corresponds to
  offset azimuth and is oriented in the direction of disk
  rotation.  \label{fig:polarflip}}
\end{figure}

The Doppler flip extends over $\sim$120\,deg or $\sim$56\,au in
azimuth, at a fixed orbital radius. This is best seen in the polar
expansion of Fig.\,\ref{fig:polarflip}, which also shows that the
intrinsic velocity deviations $\vec{u}$ must reach $\pm 1/3$
Keplerian.  As noted by \citet{CasassusPerez2019ApJ...883L..41C},
along the disk minor axis the line of sight of the azimuthal
components cancels, $u_\phi \hat{\phi} \cdot \hat{s} =0$. This implies
that $u_r$ or $u_z$ contribute significantly to
$\delta v_\circ (\vec{x})$ in the blue side of the flip, at least
along the disk minor axis. Likewise, along the disk major axis the
line of sight component of a radial component cancels,
$u_R \hat{R} \cdot \hat{s} =0$, which implies that the red side of the
flip is either due to $u_z < 0$ (receding towards the back side of the
disk), or to $u_\phi >0$, accelerating beyond Keplerian rotation at
the ascending node.

With an extension of the rotation curve to 3-D, including meridional
flows, the blue side of the flip essentially disappears from the
velocity deviation map of Fig.\,\ref{fig:kin}d, while the red side
remains. The above qualitative interpretations are thus confirmed on
average, not just along the disk axes.

Interestingly, in the face-on view of the non-axial kinematics shown
in Fig.\,\ref{fig:kinfaceon}d, the blue and red sides of the flip
align closely with the brightest continuum ridge. The velocity
deviations appear to follow the peak continuum emission. Given the
hints for an optically thick 225\,GHz continuum \citep[see ][along
with the reduction of peak amplitude $I^A_1$ and the outline of the
ring in absorption in $I^A_2$]{Perez2020ApJ...889L..24P}, the bright
ridge is likely a temperature feature rather than a local increase in
density. Thus the strongest velocity deviations coincide with the
hottest continuum emission.

In what follows we consider the origin for the observed kinematics,
taking into account the constraints on the intrinsic velocity field
$\vec{u}$. We do not consider the possibility of planet-disk
interactions driven by planets inside the cavity, because the largest
kinematic signatures of such bodies is in their immediate
vicinity. The possibility of stellar companions inside the cavity is
ruled out by the sparse-aperture-masking data presented in
\citet{Perez2020ApJ...889L..24P}.

\subsection{A protoplanet outflow}

The velocity fields traced by the rarer CO
isotopologues do not exhibit such strong velocity deviations as in
$^{12}$CO(2-1) \citep{Perez2020ApJ...889L..24P}.  The velocity
disturbance thus stems from high above the mid-plane (at the $^{12}$CO
unit-opacity surface, with an aspect ratio 2--5 times larger than
that of the thermal scale height, $h_1 \sim 2-5 h$). On the other
hand,   the blue side of the flip corresponds to a strong
vertical velocity components. Thus, the Doppler flip in the disk of HD\,100546 could perhaps
correspond to an outflow launched by a compact body at the location of
the flip, that gains in velocity with distance from the source, as in
stellar winds.

A wind in the disk of HD\,100546, approximately stemming from the location of the
flip, has in fact been proposed based on unresolved high-velocity SO
rotational line emission \citep{Booth2018A&A...611A..16B}. In addition
to the very distinct kinematics of the SO lines vs. CO(3-2),
sulfur-bearing molecules are known tracers of shocks and outflows
\citep{Booth2018A&A...611A..16B}, that release the sulfur locked in
dust grains. In this scenario, SO should be observed downstream of the
outflow, and should coincide with the blue-shifted part of the CO(2-1)
flip. The hotter continuum emission may perhaps correspond to the
shocked material.

The outflowing velocities seen in SO could be rooted in a jet launched
by planetary accretion through a circum-planetary disk, as predicted
by MHD models \citep{Gressel2013ApJ...779...59G,
  Machida2006ApJ...649L.129M}. For the flip to be consistent with the
geometry of an outflow, it is necessary that the blue side of the flip
corresponds to material whose bulk flow lies along the disk vertical
axis. If the outflow was aligned with the disk mid-plane, given the
phase of the flip its effect would be a Doppler flip opposite in
sign. The red side of the flip should correspond to material flowing
away from the observer, and therefore escaping from the back side of
the disk.

But the outflow scenario faces problems, as the continuum ring appears
to be quite optically thick.  The $^{12}$CO(2-1) channel maps seem to
outline the ring, and the back side in $^{12}$CO(2-1) cannot be seen
under the ring \citep{Perez2020ApJ...889L..24P}. In addition, the
blue side curves along a constant radius, thus following the
trajectory of bound gas in a circular orbit, rather than the bipolar
geometry of an outflow.

Finally, existing theoretical studies on protoplanet MHD polar
outflows have considered massive bodies
\citep[][]{Gressel2013ApJ...779...59G,Machida2006ApJ...649L.129M}. As
mentioned above, a giant planet should displace larger dust grains
from its orbital track. Mid-plane outflows from smaller planets in the
super-earth-mass regime have been produced by non-MHD hydro-dynamical
simulations \citep[][]{Kuwahara2019A&A...623A.179K}. Perhaps such
bodies could act like outflow sources if their equatorial plane is
inclined relative to that of the circumstellar disk. The impact of a
magnetic field in the super-earth mass regime regime remains to be
investigated.




\subsection{A disk eruption}

Perhaps the Doppler flip could be more accurately described as a {\em
  disk eruption}, a surface disturbance driven by an embedded
protoplanetary outflow. This outflow would not puncture the $^{12}$CO
layer,  and  would only push molecular material both vertically and in
the direction of rotation. In this hypothesis the planetary outflow
does not escape the disk, and instead deposits momentum in disk
material that is predominantly in Keplerian rotation, which thus
curves around in its orbit in a fashion that is reminiscent of
high-altitude volcanic plumes.





We now consider what kind of outflow would be needed to drive the
observed intrinsic velocity deviations through such a disk eruption.
We hypothesize that the observed $\delta v_\circ$ are compatible with
the flow resulting from an expansion impulse $\vec{\Gamma}$, but that
this flow does not puncture the surface. This impulse corresponds to
the force per unit volume and mass caused by the ram pressure
$\rho v_s^2$ inside an expanding shell with mass density $\rho$, width
$\Delta r_s$ and radius $r_s$,
\begin{equation}
\Gamma = ~ \frac{v_s^2}{\Delta r_s}   \gtrsim~  \frac{v_s^2}{ r_s}.
\end{equation}
A  shell with mass $M_s$   will not puncture the surface if its impulse
is balanced by the external pressure $\rho c_s^2$,
\begin{equation}
\frac{d}{dt} \left( M_s v_s \right) \sim   \rho c_s^2 4 \pi r_s^2. 
\end{equation}
We avoid the question of the launching mechanism, and consider a shell
blown at constant $v_s$ by an embedded wind or outflow,  sustained with a mass loss rate
$\dot{M}_s$:
\begin{equation}
\dot{M}_s v_s  \sim \rho c_s^2 4 \pi r_s^2. 
\end{equation}
If we evaluate $\rho$ at the CO(2-1) unit-opacity surface $H_1$ with a hydrostatic  density profile,
\begin{equation}
\frac{\dot{M}_s}{\Sigma_g} \sim \frac{r_s^2}{v_s} \frac{H_1}{f_1} e^{-\frac{1}{2} f_1^2} \Omega_K^2, 
\end{equation}
where $f_1 = H_1/H \sim 2-5$, $c_s = \Omega_K H$, and $\Sigma_g$ is
the gas surface density. For the outflow to have an appreciable
impact, we require $r_s \gtrsim H$.  At the center of the flip in
the disk of HD\,100546 ($M_\star \sim 2\,$M$_\odot$), we have a cylindrical radius
$R_f \sim 28$\,au and a disk aspect ratio given by {\tt ConeRot} of
$h = \tan(\psi) = 0.2$ for $\psi = 11$\,deg (which gives a mid-plane
temperature $T_m = 44\,$K if $f_1 = 4$, for a mean molecular weight
$ \mu= 2.17$).  If we set $v_s \sim v_f \sim 1\,$km\,s$^{-1}$, then
$\frac{\dot{M}_s}{\Sigma_g} \sim 10^{14}\,$cm$^{2}$\,s$^{-1}$.  In
other words, for $\Sigma_g \sim 5$\,g\,cm$^{-2}$, an outflow with
$\dot{M}_s \sim 10^{-6}\,{\rm M}_{\oplus}\,{\rm yr}^{-1}$ should have
an appreciable impact.

If we adopt a standard ejection to accretion mass ratio of $\sim$10
\citep[][]{Konigl2000prpl.conf..759K}, we see that the growth of a
super-earth over 1\,Myr could drive such a disk eruption at an
altitude over the disk mid-plane of 4\,$H$. It is interesting to note
that the above argument, if extended to $f_1 = H_1 / H = 5 $ (so with
mid-plane temperature of 28\,K), yields a much smaller mass loss rate
of $\dot{M}_s \sim 10^{-8}\,{\rm M}_{\oplus}\,{\rm yr}^{-1}$, which
would correspond to the accretion of a few percent of a super-earth
mass over 1\,Myr. We conclude from this argumentation that, while
embedded outflows would have little impact in the denser mid-plane,
they should lead to conspicuous velocity deviations at higher
altitudes, which is consistent with the basic properties of the
velocity deviations seen in the disk of HD\,100546.

%

\subsection{Disk mass-loading hot spot?}


Could the observed non-axial kinematics be caused by an accretion hot
spot onto the surface of the disk at the location of the flip?
\citet{Garufi2022} report localized SO and SO$_2$ line emission right
at the convergence of {\bf proposed} accretion streamers onto the
disks in HL\,Tau and DG\,Tau, where they also observe strong
non-Keplerian kinematical signatures.  HD\,100546 is a system with a
known envelope \citep{Grady2001AJ....122.3396G} that has been linked
to the idea of cloudlet-capture as a secondary (or tertiary) {\bf
  ongoing} accretion phase \citep[][]{Dullemond2019A&A...628A..20D}.
Even though there is no relevant detection of accretion streamers in
the CO(2-1) data around the region of interest, some features show
similarity to the systems where such asymmetric infall has been
proposed, as some spirals have been observed both in the optical
\citep{Grady2001AJ....122.3396G,Ardila2007} and near-infrared
\citep{Avenhaus-2014b,Sissa2018}. On top of the spiral structure seen
on large scales of several arcsec, \citet{Avenhaus-2014b} additionally
discovered a spiral in the inner system for which the base roughly
coincides with the observed kinematic deviation presented here.
\citet{Lesur2015} and \citet{Hennebelle2017} showed that asymmetric
accretion from a surrounding envelope can produce unstable accretion
shocks in the disk, from which spirals are generated and propagate
through the disk.  How the kinematics look close to the shock location
still remains an unexplored topic of interest.

We discuss the possibility of the Doppler-flip signal being linked to
such an accretion hot spot: in this scenario the red part of the
Doppler flip traces vertically infalling material that locally heats
the disk and deposits its momentum.  An accretion shock could be an
explanation for the localised detection of SO emission
\citep[][]{Booth2018A&A...611A..16B}, as thermal desorption produces
hot SO gas molecules in the post-shock area
\citep[][]{Aota2015ApJ...799..141A}. In the scenario proposed here,
the post-shock area would coincide with the blue-shifted part of the
kinematic signal; the SO detection by
\citet[][]{Booth2018A&A...611A..16B} is also strongly blue-shifted.
Whether a localised hot spot of infalling material can also be
consistent with the blue-shifted side of the signal and whether it can
dynamically perturb the dust structures at the mid-plane is an
important question that requires future numerical investigation.


\section{Conclusions} \label{sec:conc}

In this letter we presented new observations of HD\,100546 in
$^{12}$CO(2-1), and adjacent continuum, at unprecedented angular
resolutions, which we interpreted in terms of 3-D velocity deviations
around the axially symmetric flow driven by Keplerian rotation. Our
main conclusions are:

\begin{itemize}
\item The previously reported Doppler flip stands out as large
  velocity deviation from a purely tangential and axially symmetric
  flow, of $\pm1.9\,\,$km\,s$^{-1}$, corresponding to intrinsic
  velocity deviations of $\pm 1/3$ Keplerian (Fig.\,\ref{fig:polarflip})

\item However, the blue-side of the Doppler flip disappears when
  accounting for vertical and radial flows in the axially symmetric
  background flow (Figs.\,\ref{fig:kin} or \,\ref{fig:kinfaceon}
  ). Given the length of the blue side, which extends 90\,deg in
  azimuth along a fairly constant radius, the bulk of the velocity
  deviations should thus be vertical (since the height of the emitting
  surface is fixed at the $\tau=1$ surface).

\item The non-axial velocity deviations are very closely aligned with
  the brightest ridge along the continuum ring, and the center of the
  flip coincides with a bifurcation along this ridge (see
  Fig.\,\ref{fig:unsharpflip}).
  
\item The central mass is $ 2.1 < M_\star /M_\odot < 2.3$, in
  agreement with photospheric mass estimates
  (Fig.\,\ref{fig:rotcurve}).
\end{itemize}

Pending dedicated hydro-dynamical simulations, we tentatively propose
a qualitative interpretation for the Doppler flip in terms of a
surface disturbance to the Keplerian flow, i.e. a disk eruption,
driven by an embedded outflow launched by a $\sim 10\,M_{\rm earth}$
body. Such an outflow would deposit vertical momentum in disk material
that is predominantly in Keplerian rotation, and would pollute the
disk material in sulfur-bearing species downstream from the source of
the outflow, thus accounting for the SO line data.  Another possible
interpretation could involve a disk accretion hot-spot at the
convergence of a disk-mass-loading streamer infalling from the
envelope.

Whichever the nature of the strong velocity deviations in the disk of  HD\,100546,
the present data and analysis show that protoplanet searches based on
disk kinematics should consider alternative interpretations to the
non-Keplerian structures observed in molecular line channel maps. The
largest velocity deviations can be vertical or radial, and seen on the
disk surface, rather than  azimuthal and in the disk midplane, as
would be expected from the gravitational influence of an embedded and
massive planet on its immediate vicinity.


\begin{acknowledgments}
  We thank the anonymous referees for comments that improved the manuscript.
  S.C., M.C. and P.W. acknowledge support from Agencia Nacional de
  Investigaci\'on y Desarrollo de Chile (ANID) given by FONDECYT
  Regular grants 1211496, ANID PFCHA/DOCTORADO BECAS
  CHILE/2018-72190574, ANID project Data Observatory Foundation
  DO210001, ALMA-ANID postdoctoral fellowship 31180050 and FONDECYT
  Postdoctorado grant 3220399. S.P. acknowledges support from ANID-FONDECYT Regular grant 1191934. This work was funded by ANID --
  Millennium Science Initiative Program -- Center Code NCN2021\_080.
  This paper makes use of the following ALMA data: ADS/JAO.ALMA\#{\tt
    2018.1.01309.S}. ALMA is a partnership of ESO (representing its
  member states), NSF (USA) and NINS (Japan), together with NRC
  (Canada), MOST and ASIAA (Taiwan), and KASI (Republic of Korea), in
  cooperation with the Republic of Chile. The Joint ALMA Observatory
  is operated by ESO, AUI/NRAO and NAOJ. The National Radio Astronomy
  Observatory is a facility of the National Science Foundation
  operated under cooperative agreement by Associated Universities,
  Inc.
\end{acknowledgments}

%

\vspace{5mm}
\facilities{ALMA}


\software{{\sc MPolarMaps} \citep[\url{https://github.com/simoncasassus/MPolarMaps},][]{Casassus2021MNRAS.507.3789C},
  {\sc uvmem}  \citep[\url{https://github.com/miguelcarcamov/gpuvmem},][]{Carcamo2018A&C....22...16C}
}



\appendix

\section{Channel maps} \label{sec:maps}

\begin{figure*}
  \centering
  \includegraphics[width=\textwidth,height=!]{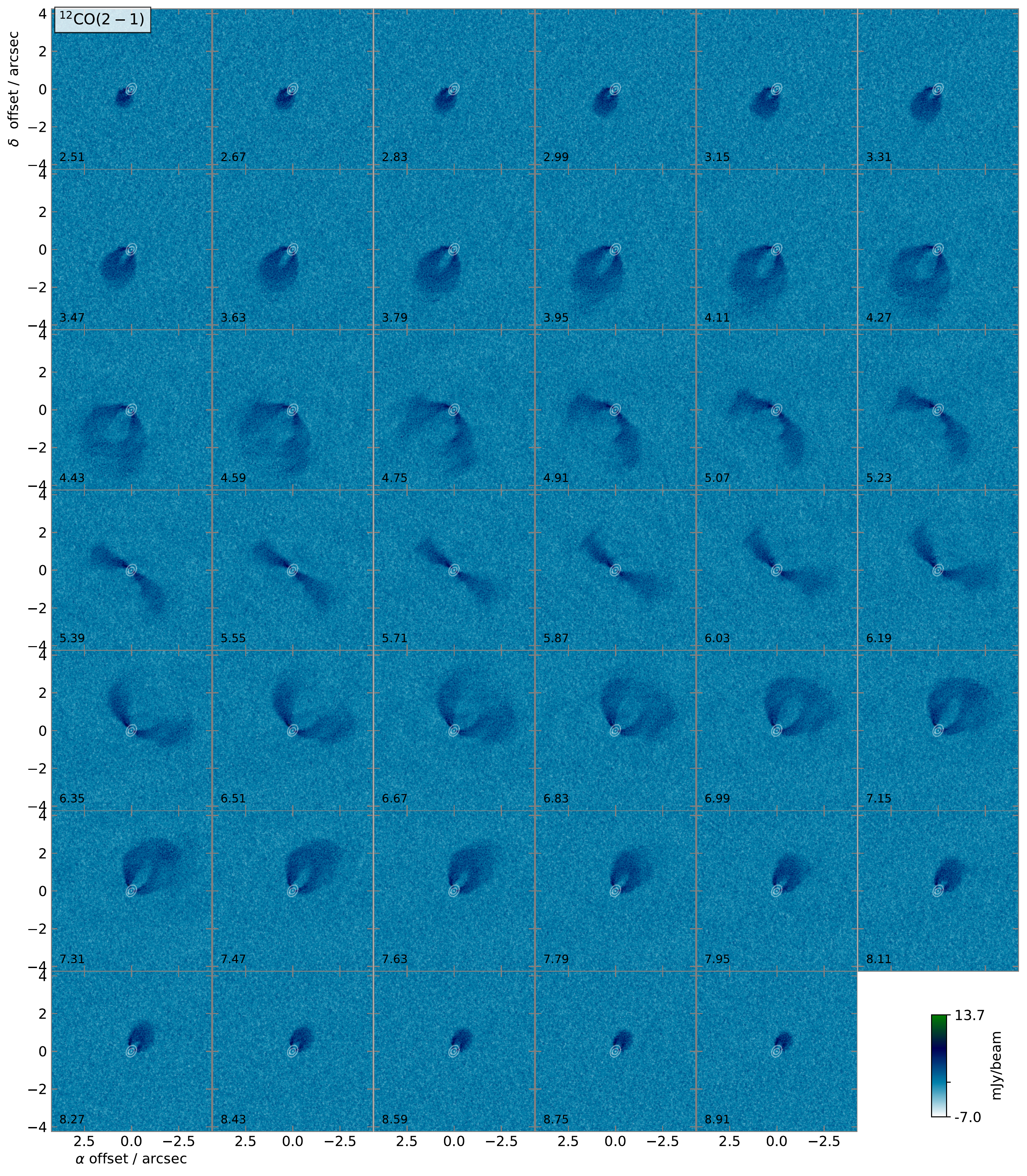}
  \caption{Channel maps from the $^{12}$CO(2-1) datacube. The
    corresponding line of sight $v_{\rm LSR}$, in km\,s$^{-1}$ is annotated  in black in
    each image.    The continuum from Fig.\,\ref{fig:cont} is outlined in contours.  The beam is $0\farcs038\times 0\farcs024 / 0.0 $, and the noise is 
    1.02\,mJy\,beam$^{-1}$.} \label{fig:channels}
\end{figure*}

The observed channels maps for the $^{12}$CO(2-1) data are shown in
Fig.\,\ref{fig:channels}.

\section{Separation of the front CO layer and moment map extraction} \label{sec:maps}

The extraction of the disk rotation curve from the velocity centroid map rests on the assumption that it corresponds to the  front CO layer, as we need to convert each line of sight to a unique location in disk-centric coordinates. As explained in Sec.\,\ref{sec:obs}, we approximately separate the two $\tau=1$ CO(2-1) layers with a double-Gaussian, where the brighter Gaussian traces the front side.  
While this approximation
breaks down in some lines of sight, such as where the back side of the
disk is exposed on the near side of the disk minor axis, a
double-Gaussian fit is in any case a better representation of the
observed profile than that obtained with a single Gaussian. In
addition, in lines of sights where the separation of the front and
back CO layers is larger than the line-width \citep[see Fig.\,3
of][for an illustration of the two CO
layers]{Pinte2018A&A...609A..47P}, the line profile is double-peaked
which for a single-Gaussian fit would bias the  velocity centroid
in-between the two velocity components.

\begin{figure*}
  \centering
  \includegraphics[width=\textwidth,height=!]{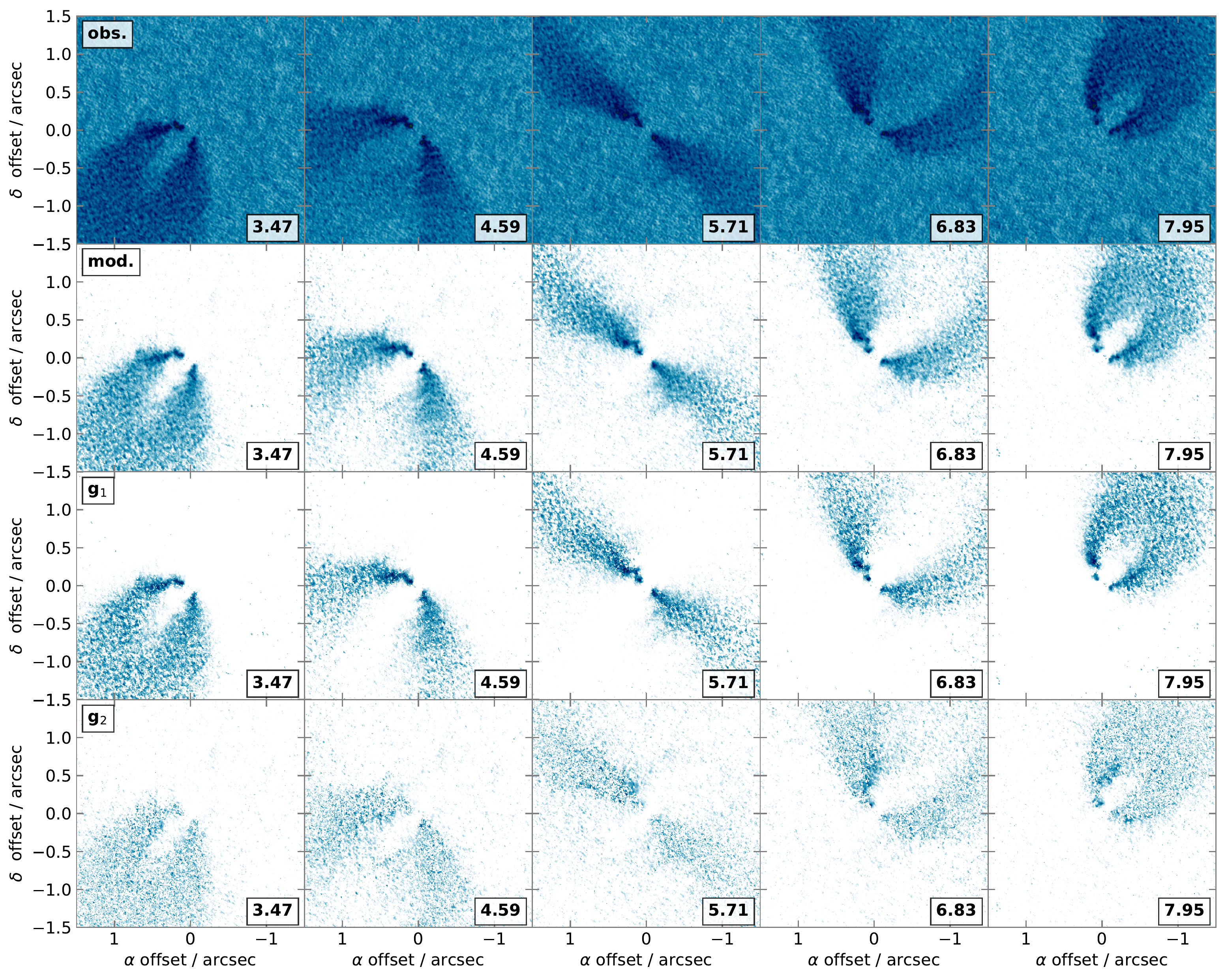}
  
  \caption{Selected channel maps derived from the double-Gaussian fits to the
    $^{12}$CO(2-1) datacube of HD\,100546 presented here. The label at
    the bottom right of each image correspond to $v_{\rm LSR}$ in
    km\,s$^{-1}$. From top to bottom, each rows shows the observed
    maps ({\bf obs}), then the double-Gaussian model image ({\bf
      mod}), then that corresponding to the brighter Gaussian ({\bf
      $g_1$}), and finally to that of the fainter Gaussian ({\bf
      $g_2$}). The two Gaussian components show conspicuous
    differences, and each roughly corresponds to the front and back
    disk surfaces.} \label{fig:2gausschans}
\end{figure*}

The separation of the two CO(2-1) layers results in qualitatively different
structures for $v^\circ_1$ and $v^\circ_2$.  The position of the peak
velocity centroid of the brighter Gaussian ($v^\circ_1$), in absolute
value, drifts aways from the disk major axis with distance from the
star, roughly towards the north-east, as expected for a flared surface
(see Fig.\,\ref{fig:moments}c). The opposite holds for the fainter
Gaussian ($v^\circ_2$), whose peak velocity drifts towards the
south-west (see Fig.\,\ref{fig:moments}f). In addition, the continuum
ring appears to coincide with a region of lower secondary Gaussian
amplitude ($I^A_2$, see Fig.\,\ref{fig:moments}e), and can be outlined
(roughly) in absorption.  The separation of the two CO(2-1) layers, using
the double-Gaussian fits, is illustrated in selected channels in
Fig.\,\ref{fig:2gausschans}.


\begin{figure*}
  \centering
  \includegraphics[width=\textwidth,height=!]{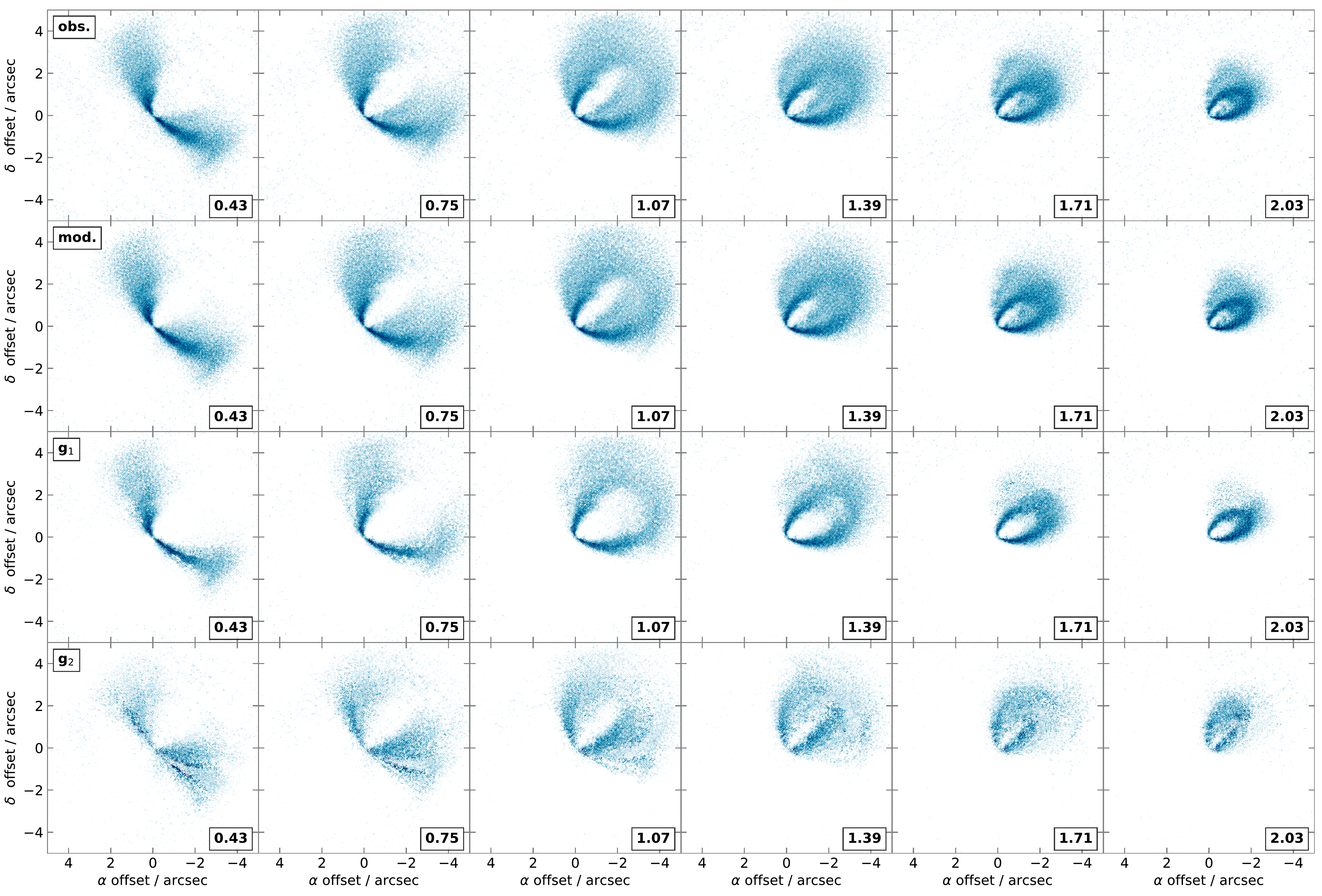}
  
  \caption{Selected channel maps obtained with  double-Gaussian fits to the
    $^{12}$CO(2-1) datacube of HD\,163296, re-imaged from  archival ALMA
    data. Annotations follow from Fig.\,\ref{fig:2gausschans}.} \label{fig:HD1632962gausschans}
\end{figure*}

\begin{figure*}
  \centering
  \includegraphics[width=\textwidth,height=!]{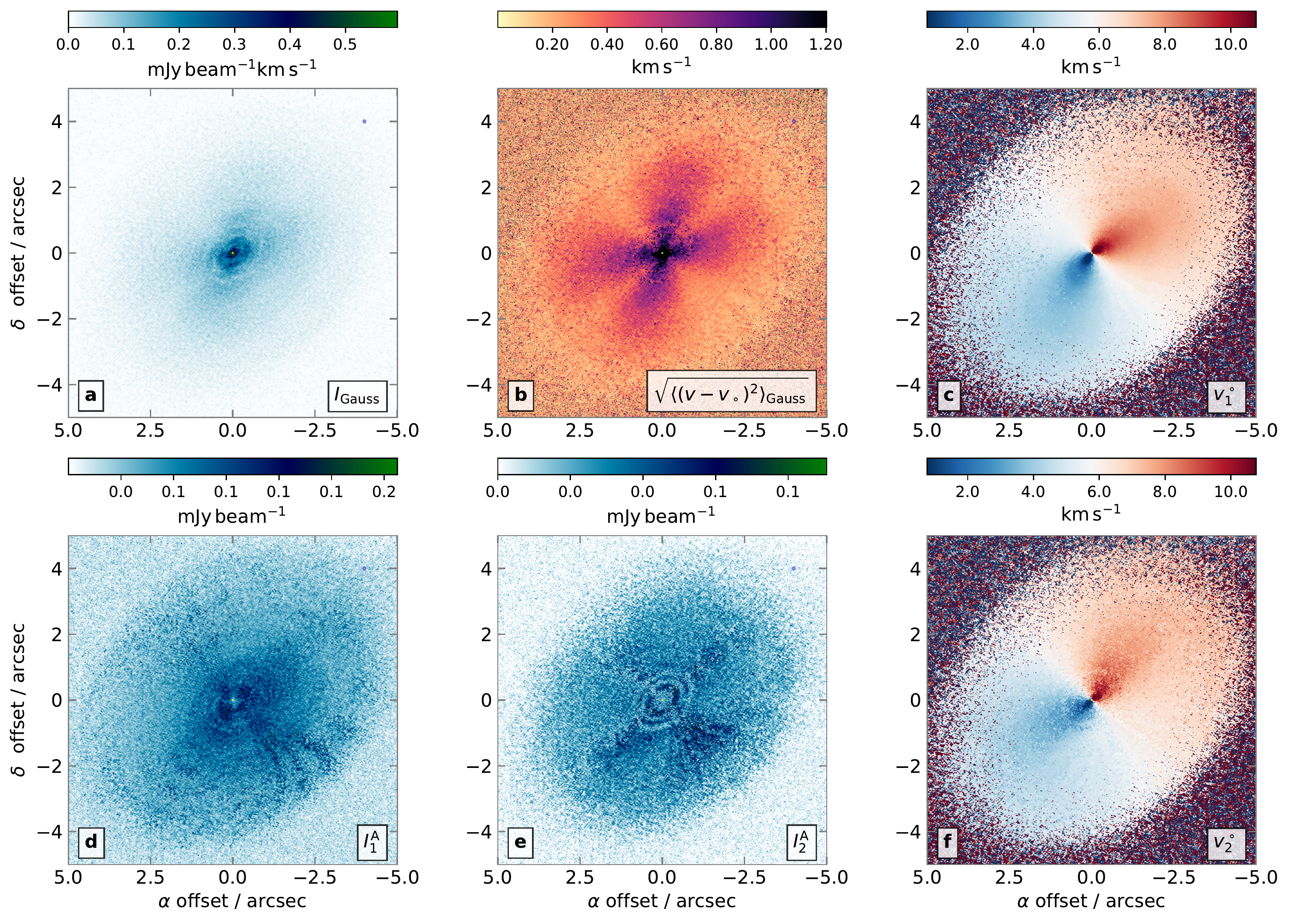}
  
  \caption{Moment maps in $^{12}$CO(2-1) towards HD\,163296, obtained
    with double-Gaussian fits in velocity. Annotations follow from
    Fig.\,\ref{fig:moments}.} \label{fig:momentsHD163296}
\end{figure*}

The continuum ring in the disk of HD\,100546 has a rather small
radius, of about 5 to 10 beams even in these long baseline data, and
the central cavity is devoid of CO(2-1) emission. In addition the
stand-alone long-baseline data of HD\,100546 are not ideal to trace
the larger scales. As a proof of concept for the use of the
double-Gaussian fit to separate the two Gaussian layers, we have 
applied {\sc GMoments} to archival data of HD\,163296 in
$^{12}$CO(2-1) \citep[from the DSHARP ALMA Large Program][, which we
re-imaged in the same way as for the HD\,100546 data presented
here]{Isella2018ApJ...869L..49I}.  The intensity maps for each CO
layer is shown for selected channels in
Fig.\,\ref{fig:HD1632962gausschans}. The corresponding moment maps are
shown in Fig.\,\ref{fig:momentsHD163296}, where the concentric
continuum ring system of HD\,163296 is seen in absorption in $I^A_2$.

\section{Extraction of 3-D rotation curves with {\sc ConeRot}} \label{sec:conerot}

We use the same notation and reference frames as in
\citet{CasassusPerez2019ApJ...883L..41C}. The sky frame is represented
by $\mathcal{S}$ and is orientated using Cartesian coordinates
$(x,y,z)$, with $\hat{y}$ aligned due North. A rotation about the
vertical axis $\hat{z}$ into frame $\mathcal{S}^\prime$ aligns
$\hat{y}\prime$ with the disc major axis, while the
$(\hat{x}\prime,\hat{y}\prime)$ plane is still parallel to the sky. A
second rotation about axis $\hat{y}^\prime$, by the inclination angle
$i$, matches the $(\hat{x}\prime,\hat{y}\prime)$ plane with the disc
midplane in frame $\mathcal{S}^{\prime\prime}$.

If all of the emission originates from the top side of the disk, which
is facing the observer, we have a bijection between the line of sight
$\vec{x}$, and the polar coordinate of its intersection with the
surface of the cone representing the disc surface, defined as the
surface of unit opacity, with height $H_1(R)$ above the midplane. This
cone has opening angle $\psi = \arctan(h_1)$, where $h_1 = H_1/R$ is
the aspect ratio of the disc surface. We thus transform the sky
coordinates $\vec{x}$ to cylindrical polar coordinates on the surface
of a cone $(R,\phi)$, as measured in $\mathcal{S}^{\prime\prime}$. The
coordinate transform from cartesian coordinates in
$\mathcal{S}^{\prime\prime}$ into cylindrical coordinates in
$\mathcal{S}^{\prime\prime}$,
\begin{equation}
  \vec{x} = \vec{f}_{\mathrm{PA},i,\psi}(R,\phi), \label{eq:f}
\end{equation}
depends on disc orientation and is
invertible. Similar coordinate transforms
have been used in, for example, \citet[][]{Rosenfeld2013ApJ...774...16R} or in 
  \cite{Isella2018ApJ...869L..49I}. We implement $\vec{f}$ with the
following formulae, already provided in in
\citet{CasassusPerez2019ApJ...883L..41C} but that we reproduce here
for completeness:
\begin{eqnarray}
    x^\prime& = &R  \sin(\phi) / \cos(i) + (H_1 - R \sin(\phi) \tan(i))\sin(i),  \label{eq:xforward}\\ 
    y^\prime& = &R \cos(\phi). \label{eq:yforward}
\end{eqnarray}
The invertion of  Eqs.~\ref{eq:xforward} and \ref{eq:yforward} to obtain
$(r,\phi) = \vec{f}^{-1}_{\mathrm{PA},i,+\psi}(x^\prime,y^\prime)$ can be achieved by 
noting that $R$ is the root of
\begin{equation}
  \frac{y^{\prime\, 2}}{R^2} + \frac{(x^{\prime}-H_1(R)\,\sin(i))^2}{R^2\cos^2(i)} = 1,
\end{equation}
and   with
\begin{equation}
 \cos(\phi) = y^\prime / R.
\end{equation}

If all of the observed signal stems from the top side of the disk, and
in the $H_1(R)$ surface, then the first moment of the axially
symmetric flow along line of sight $\hat{s}(\vec{x})$ is, using Eq.\,\ref{eq:f},
\begin{equation}
v^m_\circ(\vec{x}) = \hat{s}(\vec{x}) \cdot \vec{\tilde{v}}\left(\vec{f}_{\mathrm{PA},i,\psi}^{-1}(\vec{x})\right),
\end{equation}
or
\begin{equation}
v^m_\circ(R,\phi,z=H_1) =  \tilde{v}_\phi(R)\cos(\phi)\sin(i) + \tilde{v}_r(R) \sin(\phi) \sin(i) + \tilde{v}_z(R)\cos(i), \label{eq:vmodel}
\end{equation}
where all coordinates are measured in $\mathcal{S}^{\prime\prime}$,
i.e.  in the frame of the disc. Similar formulae for the model velocity centroid have been proposed by \citet{Teague2019Natur.574..378T}.

The origin of $\phi$ coincides with the position angle (PA) of the
disc major axis, defined as the line of ascending nodes (so the red
side). The sign convention used here results in $\tilde{v}_r(R) < 0$
for accretion. In case of prograde rotation, with $i < 90\,$deg, a
wind flowing away from the midplane corresponds to $\tilde{v}_z(R)
>0$. But for retrograde rotation, the top side of the disk that is
facing the observer corresponds to $z<0$ in
$\mathcal{S}^{\prime\prime}$, so the disk opening angle $\psi < 0$, and for a wind $\tilde{v}_z(R) < 0$.
The observed velocity deviations are thus
$\delta v_\circ (\vec{x}) = v_\circ - v^m_\circ$, and derive from the
intrinsic velocity deviations $\vec{u}(\vec{r})$
\begin{eqnarray}
  \delta v_\circ(\vec{x}) & = &   u_\phi(R,\phi)|_{z=H_1 } \cos(\phi)\sin(i)\\ \nonumber
&&  + u_R(R,\phi)|_{z=H_1 } \sin(\phi) \sin(i) \\ \nonumber
&&  +  u_z(R,\phi)|_{z=H_1 }  \cos(i). \label{eq:dvmodel}
\end{eqnarray}
 
Given the disc orientation (PA, $i$, $\psi$), we solve for the disc
rotation curve $\vec{\tilde{v}}_\circ(R)$ by performing a least
squares fit of $v^m_\circ(R,\phi)$ in Eq.\,\ref{eq:vmodel} to $v_\circ(\vec{f}_{\mathrm{PA},i,\psi}(R,\phi))$, which
is the observed velocity centroid resampled by
$\vec{f}_{\mathrm{PA},i,\psi}$.  We obtain $ \tilde{v}_\phi(R_l)$,
$\tilde{v}_r(R_l)$ and $\tilde{v}_z(R_l)$, for each discretized value
of $R_l$, by minimizing
\begin{equation}
  \chi^2_{\tilde{v}} = \sum_{k \in \mathcal{D}_\phi }w(R_l,\phi_k) ( v_{\circ}(R_l,\phi_k)- v^m_\circ(R_l,\phi_k) - v_s)^2,  \label{Eq:chi2vtilde}
\end{equation}
which assumes that the systemic velocity $v_s$ is known. In practice
we solve for the root of $\vec{\nabla} \chi^2 = 0$ using the Cramer
rule. The sum in Eq.\,\ref{Eq:chi2vtilde} runs over all azimuths
within a domain $\mathcal{D}_\phi$, with weights
\begin{equation}
w(R_l,\phi_k)=\frac{1}{N_b(R_l)\sigma^2_\circ(R_l,\phi_k)},
\end{equation}
where $N_b$ is the number of beam major axis along a circle with
radius $R_l$, clipped so that $N_b >1$. $N_b$ approximately corrects
for correlated pixels in \label{Eq:chi2vtilde}. The domain in azimuth
$\mathcal{D}_\phi$ is typically $[0,2\pi]$, but it is interesting to
restrict $\mathcal{D}_\phi$ to $[0,\pi]$, corresponding to the far
side of the disc. The model velocity centroid $v^m_{\circ}(R,\phi)$
represents the axially symmetric flow, and can be converted back to
$\mathcal{S}^{\prime\prime}$ by resampling with
$\vec{f}^{-1}_{\mathrm{PA},i,\psi}$.

If $v_s$ is not known, then we initially restrict to a purely
azimuthal rotation cuve and fit for $v_s(R_l)$ and
$\tilde{v}_\phi(R_l)$ simultaneously, for all $\{R_l\}$ in a
relatively narrow radial domain $[R_1,R_2]$, which we call a
region. This optimization also includes the disc orientation (see next
paragraph), but for constant orientation parameters ($i_\circ$,
PA$_\circ$, $\psi_\circ$) representative of this region. The value of
$v_s$ is then fixed to its median, and its uncertainty to its standard
deviation, both extracted over overlapping regions that cover a wider
range in radii. The full rotation curve is then calculated using this
global $v_s$ value.


Discs are flared, i.e.  the disc aspect ratio $h_1(R) = \tan(\psi(R))
/ R$ is a function of $R$. With molecular line tracers we expect that
$h_1(R)$ will initially increase with $R$ and then drop to zero as the
tracers becomes optically thin at the outer edge of the disc. Discs
may also warp, so that disc orientation will depend on $R$. We
calculate the disc orientation profile ($i(R), {\rm PA}(R), \psi(R)$) by
performing the optimization in overlapping radial bins.  The full
radial extension of the disc is divided into $M$ overlapping radial
bins $\{[R_{1j}, R_{2j}]\}_{j=1}^{M}$, thus defining radial regions
$\{ \Theta_j(\vec{x}^{\prime}) \}_{j=1}^{M}$, where
$\Theta_j(\vec{x}^{\prime})=1$ if $R = f_r^{-1}(\vec{x}^{\prime}) \in
[r_{1j}, r_{2j}]$, and $\Theta_j(\vec{x}^{\prime})=0$ otherwise. In
each radial bin, we maximize the log-likelihood function,
$-0.5\chi^2$ to obtain $(i_j, {\rm PA}_j, \psi_j)$ with
\begin{equation}
  \chi^2_{\rm var} = \frac{1}{\sigma_{\rm ref}^2} \sum_{l=l_1}^{l_2} \frac{ 
  \sum_{k=0}^{N_\phi-1} w(R_l,\phi_k) (v_\circ(R_l,\phi_k) -
  v^m_{\circ}(R_l,\phi_k))^2}{ \sum_{k=0}^{N_\phi-1} w(R_l,\phi_k)}, \label{eq:Chi2stddev}
\end{equation}
which is the radial sum of the weighted azimuthal variance of
residuals relative to the expected variance $\sigma_{\rm ref}^2$, with
$\sigma_{\rm ref}$ typically of order 0.1\,km\,s$^{-1}$ when including
image synthesis and channelization systematics. The  sums extend
over an interval in radius $[R_1,R_2]$ and over all azimuths.  We thus
obtain an orientation profile
$\left\{i(R_j), {\rm PA}(R_j),\psi(R_j) \right\}_{j=1}^{M}$ as well as
axially symmetric models $\{ v^m_{j}(\vec{x})\}_{j=1}^M$ in all radial
regions. The axially symmetric model corresponding to these profiles
is approximated by the average over all regions, i.e.
\begin{equation}
  v^m_{\circ R}(\vec{x}) = \frac{\sum_{j=1}^M
    v^m_{\circ}(\vec{x}) \Theta_j(\vec{x})} {\sum_{j=1}^M
    \Theta_j(\vec{x})}.
\end{equation}
This approximation ignores shadowing between regions along a line of
sight, and is thus only applicable at low to moderate inclinations.

The minimization of Eq.\,\ref{eq:Chi2stddev} and the optimal
orientation parameters, along with their associated errors, were
achieved with a Markov chain Monte Carlo ensemble sampler
\citep{MCMC2010CAMCS...5...65G}. We use the {\tt emcee} package
\citep[][]{emcee2013PASP..125..306F}, with flat priors, and with
typically 300 iterations and 30 walkers.

The azimuthal rotation curve $\tilde{v}_\phi$ can be used to constrain
the stellar mass under the assumption of pure Keplerian rotation. As
noted in \cite{CasassusPerez2019ApJ...883L..41C}, the stellar mass is
then bracketed by two extreme cases.  Assuming a rigid vertical
structure, in which the midplane velocity is the same as that at
height $H_1$, sets a lower limit. Pure vertical Keplerian shear,
ignoring radial hydrostatic support, sets an upper limit, in which
midplane azimuthal velocities are $\tilde{v}_\phi(R) (1+h_1^2)^{3/4}$.

\bibliographystyle{aasjournal}



\end{document}